\documentclass[fleqn,twoside]{article}
\usepackage[headings]{espcrc2}

\readRCS
$Id: espcrc2.tex,v 1.2 2004/02/24 11:22:11 spepping Exp $
\usepackage{graphicx}
\usepackage[figuresright]{rotating}

\newcommand{\AmS}{{\protect\the\textfont2
  A\kern-.1667em\lower.5ex\hbox{M}\kern-.125emS}}

\title{Light and heavy multiquark spectroscopy%
\thanks{Invited Talk at QCD05,  12th International QCD
Conference, July 4--9, 2005, Montpellier (France), to appear in the
Proceedings,
ed.\ S.~Narison}%
\thanks{Preprint lpsc-05-121, ArXiv:hep-ph/0601043}
}

\author{Jean-Marc Richard\address[LPSC]{Laboratoire de Physique
Subatomique et Cosmologie,\\
Universit\'e Joseph Fourier--IN2P3-CNRS,\\
53, avenue des Martyrs, 38026 Grenoble cedex, France}%
        \thanks{email: {\tt jean-marc.richard@lpsc.in2p3.fr}}}
       
\runtitle{Light and heavy multiquark spectroscopy}
\runauthor{Jean-Marc Richard}

\begin{document}

\begin{abstract}
The dynamics of multiquark binding is revisited in the light of the
recent
experimental results.
It is emphasized that some configurations mixing light and heavy
flavours are among the most favourable for stable or metastable
multiquarks. The nuclear-physics type of approach predicting the
so-called hadron--hadron molecules is compared to direct studies in
terms of
quark interaction.
\end{abstract}
\maketitle

\section{Introduction}\label{se:intro}
The  recently found or claimed new hadrons aroused a renewed interest
into the spectroscopy of rare or exotic particles
\cite{exp,Klempt:2005}. The activity  in
this field is unfortunately rather fluctuating. For instance,
experiments having ample data sets
with pions, kaons, protons and neutrons did not look very seriously
at invariant mass
plots with baryon number $B=1$ and strangeness $S=1$,
until some pentaquark event was reported elsewhere. Similarly, the
theoretical
searches in the sector of exotic hadrons follow the fashion.
Shortly after the announcement of the pentaquark peaks, several
estimates of multiquark masses and properties have been published,
which, however interesting, could have benefited from more extensive
training.

An exception is the class of calculations based on potential models,
where the know-how of few-body quantum mechanics is  used to
appreciate the difference between  mere scattering states and
possible resonances. See, e.g., Hiyama et al.\ \cite{Hiyama:2005cf}.
However, these
estimates are based on a somewhat ad-hoc  extension of the
quark--antiquark potential to situations involving more quarks and
antiquarks. It would be desirable to combine this approach with
better dynamical
ingredients.
\section{Salient experimental results}\label{se:exp}
In recent months, several new hadrons have been claimed or firmly
established, among them:
low-lying scalar and axial mesons with charm and strangeness $C=S=1$,
i.e.,  flavour content $(c\bar{s})$, the so-called
$\mathrm{D}_{s,J}^*$;
double-charm baryons, $(ccq)$, seen by the SELEX experiment;
controversial candidates for light and heavy pentaquarks;
discovery of the long-awaited missing states of charmonium $\eta_c'$
and $h_c$;
evidence for new meson resonances with hidden charm, X(3872), X(3940)
and Y(4260);
not to mention new information of light scalar
mesons,  excitations of singly-charmed baryons, etc.

It is worth stressing that the new findings are not always due to an
increase of statistics. For instance, the singlet states of
charmonium have been eventually discovered using new production and
new decay
channels.

As for pentaquarks, the superposition of positive and negative
results remains very puzzling, especially for non-experts. See, e.g.,
the review by S.~Kabana at this Conference \cite{exp}. Data were
perhaps cut off too sharply in some of the analyses leading to
tentative peaks.
Even a theorist can understand that a sequential production scheme
 would produce peaks for
any pair of final-state particles, if it is restricted to collinear
events.  The robustness of peaks when opening the angles is a crucial
test for genuine resonances.
\section{Models}\label{se:models}
These discoveries stimulated an intense theoretical activity. Old
ideas about exotics were re-examined 
and confronted to other proposed mechanisms. In particular:
\subsection{Hybrids}
Years ago, it has been suggested that the quark--antiquark
system\footnote{Hybrids baryons are also predicted}  might experience
new types of excitations \cite{Giles:1977mp}, beyond conventional
radial or orbital
excitations, and this idea is confirmed by QCD sum rules and lattice
QCD. This is very similar to the spectroscopy of 
$\mathrm{H}_2{}^+$, with a first series of ``ordinary'' states
corresponding to
the two protons moving in the lowest Born--Oppenheimer potential,  and
``exotic states'' occurring when the electron is excited. It is
extremely
plausible that at least one of the X(3872), X(3940) and Y(4260)
states is an hidden-charm hybrid $(c\bar{c}g)$. The selection rules
specific to the decay and
production of hybrids have to be checked to confirm this hypothesis
\cite{Close:2005iz}.
\subsection{Chiral dynamics}
Light-quark dynamics predicts parity partners for high-lying
excitations of hadrons, a pattern which is seemingly observed
\cite{Klempt:2005},
and also parity partners of ground states with favourable quantum
numbers. See, e.g., Ref.~\cite{Glozman:2005tq,Bardeen:2003kt} for a
discussion and further references.
If the $\mathrm{D}_{s,J}^*$ are chiral partners of the ground-state
$\{ \mathrm{D}_s,\mathrm{D}_s^*\}$, the question is whether this is a
more realistic picture of the $0^+$ and $1^+$ members of the
$(c\bar{s})$ sector, or supernumerary states, with ordinary orbital
excitations awaiting identification. This latter scenario would be
reminiscent of the situation which is observed for light scalar
mesons, with too many states as compared to a naive quark-model
counting. 
\subsection{Yukawa dynamics}
Some pioneers made the observation that nuclear forces are by no
means restricted to the nucleon--nucleon system (for refs., see,
e.g., \cite{Tornqvist:2004qy} and the review by Swanson
\cite{Klempt:2005}). For any pair of
hadrons containing light quarks, a similar interaction should be
present, and might well produce bound states, if the long-range
pion-exchange is allowed and turns out to be attractive\footnote{For
non-experts, it suffices to stress that the one-pion-exchange
potential depends on spin and isospin, and does not operate on
pseudoscalar particles. For $\mathrm{D}\mathrm{D}^*$, 
$\mathrm{D}$ flips to $\mathrm{D}^*$ and vice-versa each times a pion
is emitted or absorbed.}.  Note, however, the warning by Suzuki, that
the mass-difference between $\mathrm{D}$ and $\mathrm{D}^*$
suppresses the  effectiveness of the potential induced by pion
exchange \cite{Suzuki:2005ha}.
\subsection{Borromean binding}
If you remain sceptical, arguing that the meson--meson attraction is
presumably too weak to achieve the binding of two mesons \footnote{In
our valley of tears with $d=3$ space dimensions, a short-range
attraction call for a minimal strength to produce a quantum bound
state, unlike the more favourable cases with $d=1$ or $d=2$.}, you
should not eliminate definitely the possibility of hadron molecules
with charm. The phenomenon of  ``Borromean'' binding, more familiar
in nuclear and molecular physics, tells us that a strength about 20\%
too weak for 2-body binding is sufficient for 3-body binding. Hence a
$[\mathrm{D}\mathrm{D}^*\overline{\mathrm{D}}]$ state might exist,
with the star circulating from one constituent to the other,
following the exchanged pions between them.
This idea of Borromean molecules was applied to hadrons by Bicudo in
the context of pentaquark, and further studied by others. In this
picture, the light pentaquark, for instance, is seen as a
$(\mathrm{N},\mathrm{K},\pi)$ compound, whose none of the 2-body
subsystems is stable. See,  Ref.~\cite{Bicudo:2004pr}.
\subsection{Diquark chemistry}
The success of the quark model relies on the complicated dressing of
quarks by gluons reducing approximately to massive quarks interacting
through a confining potential.  A further simplification consists of
regrouping two quarks in a baryon to form a diquark, which, in turn,
forms a quarkonium-like structure with the third quark. This leads to
a successful phenomenology of baryon spectroscopy, baryon production
in diffractive processes, etc. 

In the good old time of baryonium, where all dreams were permitted, a
colour $\bar{3}$ or 6 diquark was imagined to rotate around its
colour-conjugate antidiquark, to form new types of meson structures.
It was not clearly demonstrated, however, how this clustering occurs
in a $(qq\bar{q}\bar{q})$ system when orbital momentum is
implemented.  The question remained unsolved, and even untouched, as
baryonium  disappeared from the tables. In contrast, Martin
\cite{Martin:1985hw}
demonstrated that in a large class of models, high-$\ell$ baryons
have a $[(qq)-q]$ structure that was postulated to explain the slope
of Regge trajectories.

The concept of diquark has been recently revisited and applied to
supernumerary scalar mesons, pentaquark states and hidden-charm
mesons that cannot be too easily accommodated as mere $(c\bar{c})$
levels \cite{Jaffe:2004ph}. A triquark $(ud\bar{s})$ was even
proposed \cite{Karliner:2003dt}. This gives
an appealing and unified picture of these hadrons. If the quark
dynamics is such that three quarks never form a $(qqq)$ structure to
leave the scene to
diquarks, three diquarks or three triquarks would presumably never
combine together. Hence,
structures like $(qq)^3$, or $(ud\bar{s})^3$ are avoided, which
otherwise would be somewhat embarrassing predictions.
Diquarks are not frozen for ever, they are effective entities at work
in a given context. Hence models based on diquarks should not be
extrapolated without care. 

Recently Maiani et al.\ described the new hidden-charm resonances in
terms of diquarks $(cq)$ or $(cs)$. See, e.g.,  \cite{Maiani:2005ug}
and refs.\ there.
\subsection{Chromomagnetic binding}
The hyperfine splitting of ground-state hadrons, such as
$\Delta-\mathrm{N}$ or $\mathrm{J}/\psi-\eta_c$ is well described by
a chromomagnetic term 
$-\sum_{i<j}C_{ij}\,\vec{\sigma}_i.\vec{\sigma}_j\,\tilde{\lambda}_i.\tilde{\lambda}_j$,
inspired from one-gluon exchange, but covering a wider range of
microscopic mechanisms.

It was  stressed that this Hamiltonian might lead to coherent
attraction, i.e., assume  in certain multiquark configurations a
value which is larger (in absolute value) that the sum of its
contributions to the  hadrons  in the threshold, hence favouring
the stability of  this multiquark  against spontaneous
dissociation. This concerns in particular the dibaryon
$\mathrm{H}(uuddss)$ \cite{Jaffe:1976yi} or the 1987-vintage
pentaquark
$\mathrm{P}(\bar{c}uuds)$ (or these obtained by permuting $u$, $d$
and $s$) \cite{Gignoux:1987cn}.
The H was desperately  searched for, and the P moderately, without
success. This was explained by unfavourable effects of
flavour-symmetry or weaker short-range correlation in multiquarks.
More extreme are the heretic attempts to substitute to the above
chromomagnetic Hamiltonian terms inspired by instanton interaction,
or spin-flavour effects.

The wave of exoticism led H\o gaasen et al.\ to revisit the
chromomagnetic Hamiltonian. When flavour symmetry is properly taken
into account, the results appear somewhat at variance with respect to
the earlier estimates. For instance, light scalar mesons with hidden
strangeness are pushed up in the spectrum, and the eigenstates with
mass of about 1\,GeV have little $\mathrm{K}\overline{\mathrm{K}}$
content.
In the $(c\bar{c}q\bar{q})$, a remarkable eigenstate appears which
almost miraculously exhibits the  mass and coupling pattern of the
$X(3872)$.

\subsection{Chromoelectric binding}
Remember the (almost) local writer:
``Il ferma la porte \`a double tour.
Malheureusement, il avait oubli\'e la fen\^etre \footnote{Alphonse
Daudet, in \textsl{La ch\`evre de Monsieur Seguin}, part of ``Les
Lettres de mon Moulin''.}'',  which can be translated here as:
If none of the previous scenarios for exotic hadrons has succeeded,
do not give up, yet. Why not try something simple and straightforward
if the most adventurous speculations have failed. A nice feature of
QCD is flavour independence. For instance, the same potential
achieves a good picture of both charmonium and bottomonium families
and gives predictions for $(b\bar{c})$. The same situation is
observed in atomic physics with the same Coulomb potential acting for
electrons, muons and antiprotons. It is noticed here that while the
equal-mass positronium molecule
$(\mathrm{e}^+,\mathrm{e}^+,\mathrm{e}^-,\mathrm{e}^-)$ is weakly
bound, the unequal mass configuration $(\mathrm{p},
\mathrm{p},\mathrm{e}^-,\mathrm{e}^-)$ is deeply bound. This suggests
that exotic mesons with two units of heavy flavour, that is to say,
$(QQ\bar{q}\bar{q})$, might exist  below the threshold for
spontaneous dissociation and hence be rather narrow. This
possibility, proposed years ago, has
received further theoretical support \cite{Gelman:2002wf}.
\section{Outlook}\label{se:outlook}
Hopefully the activity developed  after the
announcement of the pentaquark will survive its disappearance.
Understanding the origin of quark confinement is, indeed, a
fascinating program. Once models are built or basic QCD calculations
are performed to account for the main properties of quark--antiquark
and three-quarks systems, it is tempting to examine higher
configurations, or molecules made of simple hadrons. 

It often happens that states containing both heavy and light
quarks and antiquarks offer the best chances for multiquark binding.
Hence these flavour states should be preferentially  looked at.
There are already indications of unusual hadron states in the
hidden--charm sector. The double charm sector is also accessible, as
demonstrated by the detection of double-charm baryons at SELEX and
double-charmonium production at B factories.

A debate cannot be avoided between the models
based on long-range hadron dynamics  and those based
mostly on direct quark interaction.  In nuclei, the Yukawa
interaction remains at work because the short-range repulsion (and
the Fermi statistics) prevents the nucleons to merge into a single
bag. In absence of evidence for such a short-range repulsion
applicable to $\mathrm{D}\overline{\mathrm{D}}{}^*$, it is not sure 
that the long-range interaction plays a leading role. This was
already a problem with the so-called quasi-nuclear of baryonium.
Shapiro, Dover and others predicted several interesting
$\mathrm{N}\overline{\mathrm{N}}$ bound states and resonances, but
never elucidated satisfactorily how they survive short-range effects.
On the other hand, the vicinity of many states to their major decay
threshold is a clear signal that the hadron--hadron physics is
present there.

I would like to thank S.~Narison for  this beautiful conference, and
R.D.~Matheus for stimulating discussions. Due to the lack of space,
it is impossible to quote here all the interesting contributions to
this field, and I apologize for the omissions.


\begin{thebibliography}{10}
\bibitem{exp}
The experimental results are presented in several talks given at this
  Conference.
\end{thebibliography}

%
\end{document}